\documentclass[12pt]{article}
\usepackage{amsfonts}

\makeatletter
\@addtoreset{equation}{section} 
\makeatother

\newcommand{\eqn}[1]{(\ref{#1})}

\def\appendix#1
{
\addtocounter{section}{1}\setcounter{equation}{0}
 \renewcommand{\thesection}{\Alph{section}}
 \section*{
 \thesection\protect\indent \parbox[t]{16.715cm}{#1}}
 \addcontentsline{toc}{section}{Appendix \thesection\ \ \ #1}
}
\newcommand{\complex}{{\mathbb{C}}} 
\newcommand{\real}{{\mathbb{R}}} 
\newcommand{\reals}{{\mathbb{R}}} 

\def\<#1,#2>{\left\langle#1,#2\right\rangle} 

\def\hil{\mathcal{H}}
\def\uni{\mathcal{U}}

\def\nn{\nonumber}
\newcommand\opname[1]{\mathop{\mathrm{#1}}\nolimits}

\newcommand{\Tr}{\opname{Tr}}

\def\be{\begin{equation}}
\def\ee{\end{equation}}
\def\bea{\begin{eqnarray}}
\def\eea{\end{eqnarray}}

\newcommand{\del}{\partial}
\newcommand{\kmin}{$\kappa$-Minkowski}

\newcommand{\ft}{\tilde{f}}
\newcommand{\gt}{\tilde{g}}

\newcommand{\al}{\alpha}
\newcommand{\bt}{\beta}

\newcommand{\x}{\hat{x}}

\newcommand{\qs}{\left[}

\newcommand{\qd}{\right]}
\newcommand{\vq}{\vec{q}}
\newcommand{\vs}{\vec{s}}
\newcommand{\vt}{\vec{t}}
\newcommand{\vk}{\vec{k}}
\newcommand{\vl}{\vec{l}}
\newcommand{\vx}{\vec{x}}
\newcommand{\vp}{\vec{p}}

\begin{document}
\begin{titlepage}
\begin{flushright}
\baselineskip=12pt DSF--20--02\\ hep--th/0209174
\\
September2002

\end{flushright}

\begin{center}

\baselineskip=24pt

{\Large\bf Generalized Weyl Systems and $\kappa$-Minkowski space}

\baselineskip=14pt

\vspace{1cm}

{\bf Alessandra Agostini, Fedele Lizzi and Alessandro Zampini}
\\[6mm]
 {\it Dipartimento di Scienze Fisiche, Universit\`{a} di Napoli {\sl
Federico II}}\\
and \\{\it INFN, Sezione di Napoli}\\ Monte S.~Angelo, Via Cintia,
80126 Napoli, Italy\\ \ \\
{\small \tt alessandra.agostini, fedele.lizzi, alessandro.zampini~~@na.infn.it}\\[6mm]

\end{center}

\vskip 2 cm

\begin{abstract}
We introduce the notion of \emph{generalized Weyl system}, and use
it to define $*$-products which generalize the commutation
relations of Lie algebras. In particular we study in a comparative
way various $*$-products which generalize the \kmin\ commutation
relation.
\end{abstract}

\end{titlepage}

\section{Introduction}
The recent nearly explosive interest in Noncommutative
Geometry~\cite{books,ticos} has been mainly concentrated on a
noncommutative space on which the coordinates have the canonical
structure~\cite{Snyder,DFR}, their commutator being a constant:
\be [x_\mu,x_\nu]=-i\theta_{\mu\nu} \label{canoncomm} \ee This is
of course just an example of a noncommutative space, albeit a very
important one for its connections with quantum mechanics. In
general, one of the most fruitful ways to deal with such spaces is
via the definition of a deformation of the usual product among
functions on the space. This way the noncommutative space is
studied as the structure space of a deformed
$*$-algebra\footnote{The $*$ here refers to the presence of an
hermitean (complex) conjugation, and has nothing to do with the
deformed products we will introduce later on.}. The definition of
this $*$-algebra is very important, as it is the first step
towards the use of Connes' machinery for the construction of
physical theories, and the construction of field theories on
noncommutative spaces. In the deformed algebra, functions are not
multiplied with the commutative (pointwise) product, but with a
new, noncommutative, product. For example equation~\eqn{canoncomm}
becomes \be x_\mu\star x_\nu-x_\nu\star x_\mu=-i\theta_{\mu\nu}
\ee One of the ways to reproduce this relation is via the
introduction of the Moyal product~\cite{GroenewoldMoyal}: \be (f
\star g)(x) :=
\frac{1}{(\pi^{2n})\det(\theta)}\int_{\reals^{2n}}\int_{\reals^{2n}}\,d^{2n}y\,d^{2n}z
e^{2i(x^t\theta^{-1} y + y^t\theta^{-1} z + z^t\theta^{-1}
x)}\,f(y)g(z)  \label{eq:Moyal} \ee This product is probably more
familiar with the asymptotic expansion \be (f\star
g)(x)\equiv\left.e^{-\frac
i2\theta_{\mu\nu}\partial^{y_\mu}\partial^{z_\nu}}
f(y)g(z)\right|_{y=z=x} \label{defstar} \ee which however is valid
on a smaller domain than~\eqn{eq:Moyal}. In terms of this product
the exponentiated version of~\eqn{canoncomm} reads: \be e^{ik^\mu
x_\mu} \star e^{il^\nu x_\nu}=e^{\frac
i2k^\mu\theta_{\mu\nu}l^\nu}e^{i(k+l)^\mu x_\mu}
\label{expcancomm} \ee Generalized deformed $*$-products were
originally introduced in~\cite{Bayenetal} as a first attempt to
develop a quantization of classical dynamics on phase space. In
fact, such a deformed product can be defined on a vector space
equipped with a constant symplectic structure. A slightly more
general case is the introduction of a deformed product on a vector
space on which a constant, at first, Poisson bracket is defined.
It has the property that the $*$-commutator of two functions
reduces to the Poisson bracket to first order in the deformation
parameter: \be [f,g]_*=-i\theta\{f,g\}+O(\theta^3) \ee A manifold
on which a Poisson bracket has been defined is called a Poisson
Manifold. All symplectic manifolds, on which a nonsingular
two-form $\omega$ is defined, are naturally Poisson manifolds,
defined in terms of the bivector $\Lambda$, inverse of $-\omega$:
\be \{f,g\}=\Lambda_{\mu\nu}\del^\mu f\del^\nu g \ee Not all
Poisson manifolds are however symplectic. The general problem of
finding a deformed product for a general Poisson manifold has been
solved by Kontsevich~\cite{Maxim}, at least at the level of formal
series, that is, without considerations of the convergence of the
series.

A noncommutative space is \kmin, defined by: \bea
{}[x_0,x_i]&=&\frac i\kappa x_i=i\lambda x_i\nn\\ {}[x_i,x_j]&=&0
\label{kmincomm} \eea The importance of \kmin\ (and the origin of
the name\footnote{The notation with the parameter
$\lambda=1/\kappa$ is also common, and we prefer it in this
paper.}) lies in the fact that it is the homogeneous space for the
quantum deformation of the $D=4$ Poincar\'e algebra called
$U_{\kappa}(\mathcal{P}_4)$~\cite{zakrz}. This deformation of the
Poincar\'e algebra has been originally presented in~\cite{LNRT} by
contraction\footnote{For an alternative construction see, for
example~\cite{delolmo}.} of a real form of the quantum anti-de
Sitter algebra $U_q(so(3,2))$ with a procedure introduced
by~\cite{firenzegroup}. The choice of the generators of
$U_{\kappa}(\mathcal{P}_4)$ is not unique, different basic
generators modify its form. Expressing its generators in the
bicrossproduct basis~\cite{MajidRuegg}, it is possible to see that
$\kappa$-Poincar\'e acts covariantly as an Hopf algebra on a
noncommutative space whose generators obey
equation~\eqn{kmincomm}. In the limit $\kappa\to\infty$
($\lambda\to 0$) one recovers the standard Minkowski space, with
the ordinary Poincar\'e group. \kmin\ is naturally a Poisson
manifold, with bracket: \be \{f,g\}=x_i(\del^i f \del^0 g-\del^0f
\del^i g) \label{poissonkmin} \ee

 A crucial aspect of these
relations is that they define a solvable Lie algebra on the
generators, and this makes them particular cases of more general
products.

The aim of this paper is to compare different products present in
the literature, which satisfy ~\eqn{kmincomm} and show that they
can be seen as generalizations of the Moyal product, or rather of
the exponentiated version of eq.~\eqn{expcancomm}, based on a more
general notion of Weyl system. Along the way we will develop a
method for the construction of deformed products, that enable to
obtain a wider class of commutation relations among the coordinate
functions, reproducing Lie algebra structure with the possibility
of considering central extension as well.

The paper is organized as follows. In section 2 we introduce the
main tool of our investigation: Weyl systems, and generalize them.
In section 3 we review a number of deformed \kmin\ products, and
in section 4 we show that they can all be written in terms of
generalized Weyl systems.  Some conclusions follow. Technical
details of the calculations are in the appendix.

\section{Weyl System and maps and their Generalization}
\subsection{Standard Weyl Systems and Maps}
The concept of what can now be referred to as a standard Weyl
system was introduced by H.~Weyl~\cite{weyl}. The main motivation
at the time was to avoid the presence of unbounded operators in
the quantum mechanical formalism. We will use the concept (and its
generalizations presented below) as a tool to construct
$*$-algebras whose generators satisfy some commutation relations.

We start with a brief description of the standard Weyl
systems~\cite{BaezSegal}. Given a real, finite dimensional,
symplectic vector space $S$, a \emph{Weyl system} is a map between
this space and the set of unitary operators on a suitable Hilbert
space:
\be
W\,:\,S\,\mapsto \,\uni\left(\hil\right)
\ee
with the property:
\be
W(k+k^{\prime})=e^{-\frac{i}{2}\omega\left(k,k^{\prime}\right)}
W(k)W\left(k^{\prime}\right) \label{stweylsys}
\ee
where $\omega$ is the symplectic, translationally invariant, form
on $S$. On each one-dimensional subspace of $S$, this formula
reduces to ($\alpha$ and $\beta$ real scalars):
\be
W\left(\left(\alpha+\beta\right)k\right)=W\left(\alpha k\right)
 W\left(\beta k\right)=W\left(\beta k\right)W\left(\alpha k\right)
\ee
This means that, for each $k$, $W\left(\alpha k\right)$ is a one
parameter group of unitary operators. According to Stone's theorem
$W$ is the exponential of a hermitian operator on $\hil$:
\be W\left(\alpha k\right)=e^{i\alpha X\left(k\right)}
\label{defgenstan} \ee and the vector space structure implies that
\be X\left(\alpha k\right)=\alpha X\left(k\right)\ee Relation
~\eqn{stweylsys} can be cast in the form \be
W(k)W\left(k^{\prime}\right)= e^{i\omega\left(k,k^{\prime}\right)}
W\left(k^{\prime}\right)W(k)
\ee
This can be considered the exponentiated version of the
commutation relations, thus satisfying the original Weyl
motivation. The usual form of the commutation relations between
generators can be recovered with a series expansion
\be
\left[X\left(k\right),X\left(k^{\prime}\right)\right]
=-i\omega\left(k,k^{\prime}\right) \ee

In the usual identification of $S$ with $\mathbb{R}^{2n}$, the
cotangent bundle of $\mathbb{R}^{n}$, with canonical coordinates
$\left(q^{i},p_{j}\right)$ and $\omega=dq^{j}\wedge dp_{j}$, this
construction can be given an explicit realization  \be
W\left(q,p\right)=e^{i\left(q^{j}P_{j}+p_{j}Q^{j}\right)}
\label{darbouxstanweylsys} \ee where $P_{j}$ and $Q^{j}$ are the
usual operators that represents the position and momentum
observables for a system of particles, whose dynamics is
classically described on the phase space $\mathbb{R}^{2n}$. This
form of the operator $W$ suggests how to relate an operator on
$\hil$ to a function defined on $S$. It reminds the integral
kernel used to define the Fourier transform. It can be intuitively
seen as a sort of "plane wave basis" in a set of
operators\footnote{Our considerations are valid for rapid descent
Schwarzian functions, and in the following we will not pay
particular attention to the domain of definition of the product,
discussed at length in~\cite{graciavarillyjmp}.}. Given a function
$f$ on the phase space, whose coordinates we collectively indicate
with $x$, with Fourier transform \be \tilde f(k)=\frac
1{(2\pi)^n}\int d^{2n}x f(x) e^{-ikx} \ee we define the operator
$\Omega(f)$ via the Weyl map \be \Omega(f)=\frac 1{(2\pi)^n}\int
d^{2n}k \tilde{f}\left(k\right)W\left(k\right)
\label{standardweyl} \ee Where by $kx$ we mean $k^\mu x_\mu$.
$\Omega\left(f\right)$ is the operator that, in the $W(k)$ basis,
has coefficients given by the Fourier transform of $f$. The
inverse of the Weyl map, also called the Wigner map, maps an
operator $F$ into a function, whose Fourier transform is: \be
\Omega^{-1}(F)(k)= \Tr F W^\dagger(k)
\ee The bijection $\Omega$ can now be used to translate the
composition law in the set of operators on $\hil$ into an
associative, non abelian, composition law in the space of function
defined on $\real^{2n}$. For two functions on $\real^{2n}$ we
define the $\star$ product as \be (f\star
g)=\Omega^{-1}\left(\Omega(f)\Omega(g)\right)
\label{stardefweylmap} \ee For a Weyl system defined
by~\eqn{stweylsys}, with $\omega$ such that
$\omega\left(k,k^{\prime}\right)=k^{\mu}\theta_{\mu\nu}k^{\prime\nu}$,
this reduces to the product defined in~\eqn{eq:Moyal}
or~\eqn{defstar}. If we consider to perform Fourier transform in a
distributional sense, then it is possible to define Moyal product
between coordinate functions, thus obtaining
\be
x_{\mu}\star x_{\nu}=x_{\mu}x_{\nu}-\frac{i}{2}\theta_{\mu\nu}
\label{coostar}\ee and relation~\eqn{canoncomm}.

The deformed algebra just defined is a $*$-algebra\footnote{We
leave aside issues of norm completeness of the algebra.} with norm
\be
\parallel f \parallel=\sup_{g\neq
0}\frac{\|f\star g\|_2}{\|g\|_2} \label{defnorm}
\ee
with $\|\cdot\|_2$ the $L^2$ norm defined as
\be
\left(\|f\|_{2}\right)^{2}\equiv \int d^{2n}k |\tilde f(k)|^2 \ee
The hermitean conjugation is the usual complex conjugation. Note
that it results
\be
\Omega(f^*)=\Omega(f)^\dagger
\ee

These two ingredients enable to give this set of functions a very
important structure in the context of non commutative geometry
formalism. It is possible to see that Weyl map $\Omega$ is an
example of the GNS construction (see~\cite{ticos}) which
represents any $C^{*}-$algebra as bounded operators on a Hilbert
space.

\subsection{Generalized Weyl Systems\label{genweyl}}
In the last section we have shown how the Moyal product arises via
an explicit realization of the Weyl system, in terms of unitary
operators on a Hilbert space, and how it is nothing but a
realization, in the space of functions, of the operator
composition law. Now we show how it is possible to define a class
of "deformed" products in a set of function defined on
$\real^{n}$, without using an explicit realization of a Weyl
system, thus opening the possibility for a generalization of this
concept.

While we have previously considered a standard Weyl system simply
as a map, now we want to consider it as a unitary projective
representation of the translations group in an even dimensional
real vector space. The most natural generalization is to consider
the manifold $\real^{n}$, and $\oplus$, a non-abelian composition
law between points.  Thus $\real^n$ acquires a general Lie group
structure. A \emph{generalized Weyl system} is the map in the set
of operators $W(k)$ with the composition rule
\be W(k)W(k')=e^{\frac{i}2\omega(k,k')}W(k\oplus k')
\label{genweylrel} \ee For the algebra to be associative it must
be \be \omega\left(k^1,k^2\oplus
k^3\right)+\omega\left(k^2,k^3\right)
=\omega\left(k^1,k^2\right)+\omega\left(k^1\oplus k^2,k^3\right)
\label{cocycle} \ee Without entering into a cohomological
characterization of this relation, it is enough to mention that
such a $\omega$ is called a cocycle. If
$\omega\left(k,k^{\prime}\right)$ is bilinear in both its entries,
then it is necessary a cocycle. In the Moyal case $\omega$ is
two-form and therefore bilinear in both entries.

Since we are looking for deformations of the algebra driven by a
parameter $\lambda$, we also require that
\bea
\lim_{\lambda\to 0}\, k\oplus k' &=& k+k'\nn\\
\lim_{\lambda\to 0}\, \omega(k,k') &=& 0 \eea In analogy with the
Moyal case we define a map $\Omega$ from ordinary functions to
formal elements of a noncommutative algebra as\footnote{We will
denote the elements of the deformed algebra with capital letters.}
\be \Omega(f)\equiv F \equiv\frac{1}{(2\pi)^{n/2}} \int d^{n}k
\ft(k) W(k)\label{defOmega} \ee where the product is \be
\Omega(f)\Omega(g)=FG=\frac{1}{(2\pi)^{n}}\int d^nkd^nk'
\ft(k)\gt(k')W(k)W(k')
\ee
Formally using the definition of generalized Weyl systems we
obtain
\be
FG=\frac{1}{(2\pi)^{n}}\int d^nkd^nk'
\ft(k)\gt(k')e^{\frac{i}{2}\omega(k,k')}W(k\oplus k')
\label{FGfourier} \ee It is useful to write the product as a
twisted convolution. In order to do this, let us define the
inverse of $k$ with respect to the composition law
$\oplus$\footnote{Here $0$ denotes the neutral element of
$(\real^{n},\oplus).$}: \be \bar k \ : \ k\oplus \bar k = \bar k
\oplus k=0 \ee Note that we are assuming that the right and the
left inverse are the same. With a change of variables in
eq.~\eqn{FGfourier} \be k\oplus k'=\xi \ee then \be
k=\xi\oplus\bar{k'}=\al(\xi,\bar{k'}) \label{trasfkxi} \ee
Equation~\eqn{FGfourier} becomes \bea
FG&=&\frac{1}{(2\pi)^{n}}\int d^nk'd^n\xi J(\xi,k')
\ft(\xi\oplus\bar{k'})\gt(k')e^{\frac{i}{2}\omega(\xi\oplus\bar{k'},k')}W(\xi)\nn\\
&=&\frac{1}{(2\pi)^{n}}\int d^n\xi
d^nk'J(\xi,k')\ft(\al(\xi,\bar{k'}))\gt(k')e^{\frac{i}{2}
\omega(\al,k')}W(\xi) \label{twistedconv} \eea where
$J(\xi,k')=|\partial_{\xi}\al(\xi,\bar{k'})|$ is the Jacobian of
the transformation~\eqn{trasfkxi}. The last equation can be cast
in a more suggestive form
\be
FG=\frac{1}{\left(2\pi\right)^{n/2}}\int d^{n}\xi
\left(\frac{1}{\left(2\pi\right)^{n/2}}\int d^{n}k^{\prime}
J(\xi,k')\ft(\al(\xi,\bar{k'}))\gt(k')e^{\frac{i}{2}
\omega(\al,k')}\right)W(\xi) \ee Comparison with~\eqn{defOmega}
suggests to define the Fourier transform of the deformed product
of $f$ and $g$ as: \be
\widetilde{\left({f*g}\right)}\left(\xi\right)=
\frac{1}{\left(2\pi\right)^{n/2}}\int d^{n}k^{\prime}
J(\xi,k')\ft(\al(\xi,\bar{k'}))\gt(k')e^{\frac{i}{2}
\omega(\al,k')}\ee A detailed calculation shows that \be
(f*g)(x)=\frac{1}{(2\pi)^{n}}\int d^nkd^nk'
\ft(k)\gt(k')e^{\frac{i}{2}\omega(k,k')}e^{i(k\oplus k')x}
\label{deformedproduct}\ee This definition enables to write
$\Omega(e^{ikx})=W(k)$, and \be \Omega^{-1}(W(k))=e^{ikx} \ee
which gives the inverse of the map~\eqn{defOmega} for all
operators that can be ``expanded'' in the ``plane wave'' basis
given by the $W$'s. As we claimed, it has been possible to define
a product simply considering $W$ as a formal device. Now we have
given the set of function on $\real^{n}$ a structure of algebra.
It is easy to check that the neutral element of the product is
$W(0)$, which, as is usual for noncompact geometries, does not
belong to the algebra, which is composed of functions which vanish
at infinity; while associativity is a consequence of the
associativity of $\oplus$.

 To define an hermitean conjugation, we use the fact that for the
undeformed algebra \be f(x)^\dagger=f(x)^*=\frac
{1}{(2\pi)^{n/2}}\int d^nk \ft^*(k) e^{-ikx}=\frac
{1}{(2\pi)^{n/2}}\int d^nk \ft^*(-k) e^{ikx}\label{conjuga} \ee
and define the hermitean conjugate of $F$, defined as in
eq.~\eqn{defOmega} as \be F^{\dag}=\frac{1}{(2\pi)^{n/2}}\int d^nk
\ft^*(\bar k)W(k) \ee This means that we assume, in the set of
functions
\be
f^{\dagger}\left(x\right)=\frac{1}{\left(2\pi\right)^{n}}\int
d^{n}k\,d^{n}a\,f^{*}\left(a\right)e^{ikx}e^{i\bar{k}a}\ee

The norm is defined as in~\eqn{defnorm}. The compatibility of the
hermitean conjugation with the product~\eqn{FGfourier}: \be
(FG)^\dagger=G^\dagger F^\dagger \ee imposes further restrictions
on $\omega$ and $\oplus$. Using the definition of hermitean
conjugate we obtain: \be (FG)^{\dag}=\frac{1}{(2\pi)^{n}}\int
d^n\xi d^nk'J^*(\xi',k')|_{\xi'
=\bar{\xi}}\ft^*(\al(\bar{\xi},\bar{k'}))
\gt^*(k')e^{-\frac{i}{2}\omega^*(\al(\bar{\xi},\bar{k'}),k')}W(\xi)\label{1}
\ee on the other hand if we compute \bea
G^{\dag}F^{\dag}&=&\frac{1}{(2\pi)^{n}}\int d^nk d^np
\gt^*(\bar{k})\ft^*(\bar{p})W(k)W(p)\nn\\
&=&\frac{1}{(2\pi)^{n}}\int d^nk' d^n\xi
J'(\xi,k')\gt^*(k')\ft^*(\overline{(k'\oplus\xi)})e^{\frac{i}{2}
\omega(\bar{k'},k'\oplus\xi)}W(\xi)\nn\\
&=&\frac{1}{(2\pi)^{n}}\int d^nk' d^n\xi J'(\xi,k')
\gt^*(k')\ft^*(\al(\bar{\xi},\bar{k'}))e^{\frac{i}{2}\omega(\bar{k'},\al(k',\xi))}W(\xi)
\label{2} \eea where $k'=\bar {k}$ and  $k\oplus p=\xi$, then \be
p=\bar{k}\oplus\xi=k'\oplus\xi\Rightarrow
\bar{p}=\bar{\xi}\oplus\bar{k'} \ee if and only if \be
\overline{(k\oplus k')}=\bar{k'}\oplus\bar{k} \label{ominus} \ee
This is a further requirement on $\oplus$, that makes it into a
group. The Jacobian becomes \be
J'(\xi,k')=|\partial_{k'}\bar{k'}\,\partial_{\xi}\al(k',\xi)| \ee
Comparing the two equations~(\ref{1}) and~(\ref{2}), we obtain
sufficient conditions for compatibility \bea
J'(\xi,k')\equiv|\partial_{k'}\bar{k'}\partial_{\xi}\al(k',\xi)|
&=&|\partial_{\xi'}\al^*(\xi',\bar{k'})|_{\xi'=\bar{\xi}}\equiv J^*(\xi',k')|_{\xi'=\bar{\xi}}\label{3} \\
\omega^*(\al(\bar{\xi},\bar{k'}),k')&=&-\omega(\bar{k'},\al(k',
\xi))\label{4} \eea The standard Weyl--Moyal system described in
the previous section is an example of this construction, with
$\oplus$ the usual sum. Once an abstract $*$-algebra has been
defined, it is in principle possible to construct an Hilbert space
on which the algebra is represented by bounded operators. This
could eventually be done with the GNS construction, enabling to
recover $W$ as explicitly realized operators.

As we said, the $\oplus$ has given a group structure to
``momentum'' space, and of course there will be a Lie algebra
associated to the group. We now will argue that this Lie algebra
structure is the same of the noncommutativity of the $x$'s on the
deformed space. Define first of all the generators $x_i$'s
\be
X_{\alpha}=\frac{1}{\left(2\pi\right)^{n}}\int
d^{n}x\,d^{n}k\,x_{\alpha}e^{-ikx}W\left(k\right) \ee Product
between them is (performing the integral in a distributional
sense, with suitable boundary conditions)\bea {}
x_{\alpha}*x_{\beta}&=&\frac{1}{\left(2\pi\right)^{2n}}\int
d^{n}z\,d^{n}y\,d^{n}k\,d^{n}l\,z_{\alpha}y_{\beta}e^{-i\left(kz+ly\right)}
e^{\frac{i}{2}\omega\left(k,l\right)}e^{ix\left(k\oplus l\right)}
\nonumber\\
&=&-\frac{i}{2}\left(\frac{\del^{2}\omega\left(k,l\right)}{\del
k_{\alpha}\del l_{\beta}}\right)\Bigg|_{k=l=0}
-ix_{\mu}\left(\frac{\del^{2}}{\del k_{\alpha}\del
l_{\beta}}\left(k\oplus l\right)^{\mu}\right)
\Bigg|_{k=l=0}+x_{\alpha}x_{\beta}\eea The commutator is given by
the antisymmetric combination of this product. It is possible to
prove that the second term in the r.h.s. of this relation gives
the structure constants of the Lie algebra defined by the Lie
group $\left(\real^{n},\oplus\right)$:
\be
\left(\frac{\del^{2}}{\del k_{\alpha}\del l_{\beta}}\left(k\oplus
l\right)^{\mu}\right)-\left(\frac{\del^{2}}{\del k_{\beta}\del
l_{\alpha}}\left(k\oplus l\right)^{\mu}\right)
\Bigg|_{k=l=0}=c^{\mu}_{\alpha\beta}\ee while the cocycle term
gives a central extension that can be cast in the usual form:\be
\frac{1}{2}\left(\frac{\del^{2}\omega\left(k,l\right)}{\del
k_{\alpha}\del
l_{\beta}}-\frac{\del^{2}\omega\left(k,l\right)}{\del
k_{\beta}\del
l_{\alpha}}\right)\Bigg|_{k=l=0}=\theta_{\alpha\beta} \ee Finally
one obtains:
\be
[x_\alpha,x_\beta]_*=-ic_{\alpha\beta}^\mu
x_\mu-i\theta_{\alpha\beta} \ee In the following we will consider
\kmin\ without central extensions, and therefore set $\theta=0$.

\section{$*$-products on \kmin}
So far we have abstractly defined a generalization of a Weyl
system. In this section we will show that these systems are a good
description of some $*$-products in the study of the
$\kappa$-Minkowski space which we briefly describe in this
section. The noncommutative space \kmin\ is intimately related to
the $\kappa$-deformed Poin{\-}ca{\-}r\'e algebra,
$U_{\kappa}(\mathcal{P}_4)$. This is a quantum group originally
obtained in~\cite{LNRT}, by contraction of the $q$-deformed
anti-de Sitter algebra in the so--called standard basis.
Subsequently in~\cite{MajidRuegg} a new basis was found
(bicrossproduct basis) in which the Lorentz sector is not
deformed, the deformation occurs only in the cross-relations
between the Lorentz and translational sectors\footnote{We will
usually use four-dimensional greek indices ($\mu,\nu=0,\ldots,3$)
and three dimensional latin indices ($i,j=1,2,3$), for \kmin\ we
use a $(+,-,-,-)$ signature. Many of our considerations are valid
in an arbitrary number of dimensions, higher dimensional versions
of $\kappa$-Poincar\'e are in~\cite{LukierskiRuegg}. }:
\bea
{}[P_\mu,P_\nu]&=&0\nn\\ {}[M_i,P_j]&=&i\epsilon_{ijk}P_k
\nn\\{}[M_i,P_0]&=&0\nn\\
{}[N_i,P_j]&=&-i\delta_{ij}\left(\frac{1}{2\lambda}(1-e^{2\lambda
P_0})+\frac{\lambda}{2} P^2\right)+i\lambda P_iP_j\nn\\
\qs N_i,P_0\qd &=&iP_i \label{kappapoincare}
\eea
and the Lorentz subalgebra remains classical:
\bea
[M_i,M_j]&=&i\epsilon_{ijk}M_k\nn\\
\qs M_i,N_j\qd &=&i\epsilon_{ijk}N_k\nn\\
\qs N_i,N_j\qd &=&-i\epsilon_{ijk}M_k
\eea
All these commutation relations becomes the standard ones for
$\lambda\to 0$. The bicrossproduct basis is peculiar as
$\kappa$-Poincar\'e acts \emph{covariantly} on a space that is
necessarily deformed and noncommutative. This is a consequence of
the non cocommutativity of the coproduct which, always in the
bicrossproduct basis, reads:
\bea
\Delta P_0&=& P_0\otimes 1 + 1 \otimes P_0\nn\\
 \Delta M_i&=& M_i\otimes 1 +1 \otimes M_i\nn\\
 \Delta P_i&=&P_i\otimes 1+e^{\lambda P_0}\otimes P_i\nn\\
  \Delta N_i
&=&  N_i\otimes 1+e^{+\lambda P_0}\otimes
N_i+\lambda\varepsilon_{ijk}P_j\otimes M_k \label{coproducts}
\eea
The quantum algebra~\eqn{kappapoincare} contains a translation
subalgebra, and it is natural to consider the dual of the
enveloping algebra of translations as  \kmin\ space, on which
$\kappa$-Poincar\'e acts covariantly. Because of the non
cocommutative relations~\eqn{coproducts} the generators of the
dual space do not commute, and their commutation relations is
given by~\eqn{kmincomm}. The possibility that at high energies the
symmetry of quantum gravity may be of the quantum kind has created
some interest in the study of \kmin\ spaces. Deformations of the
equations of motion in  momentum space for free classical fields
have been investigated in~\cite{LRR,GGKMC,M,NST}. The differential
calculus and Dirac equation have been studied
in~\cite{Bibikov,KMLS}. A field theory has been investigated
in~\cite{KLM1,KLM2,AmelinoArzano}, and the modification of
dispersion relations due to noncommutativity has been related to
astrophysical phenomena in~\cite{AmelinoMajid}. It was argued
in~\cite{SDR} that the $\kappa$-Poincar\'e Hopf algebra that
characterizes the symmetries of \kmin\ can be used to introduce
the Planck length ($\simeq \lambda$) directly at the level of the
relativity postulates (Doubly Special Relativity).

For the construction of a gauge theory, a $*$-product is a key
tool~\cite{SW,Monaco,Szabo}, and $*$ products in the context of
\kmin\ generalizing~\eqn{kmincomm} have been presented
in~\cite{KLM1,KLM2}. On the other side the commutation relations
of \kmin\ realize a Lie algebra, and hence the manifold on which
they are defined is a Poisson manifold, which has been the subject
of intense studies. They culminate with the proof~\cite{Maxim}
that it is always possible to define a $*$-product on these
manifolds, albeit in the space of formal series without warranty
of their convergence. In the following we will consider some
$*$-products which define various versions of \kmin, and show how
they can all be considered generalized Weyl systems. We warn the
reader that however our list is not complete. Notably the product
defined in~\cite{Maxim} will not be considered in this paper. We
hope to return to this problem, as well as to the problem of the
equivalence of the products, in the future.
\subsection{The CBH product}
The first product we present is a simple application of the well
known Campbell--Baker--Hausdorff (CBH) formula for the product of
the exponential of noncommuting quantities. This $*$-product is
actually a particular case of the general product on the cotangent
space of a Lie algebra, called the Gutt product~\cite{Gutt}. It
has also been investigated in the context of deformation
quantization in ~\cite{kathoia,9903036,Monaco,0006246,KLM2}. Let
us present this product in a more general way for a set of
operators satisfying a Lie algebra condition:
\be
[\x_\mu,\x_\nu]=c_{\mu\nu}^\rho\x_\rho
\ee
The CBH formula is:
\be
e^{i\al^\mu\x_\mu}e^{i\bt^\nu\x_\nu}=e^{i\gamma(\al,\bt)^\nu\x_\nu}
\label{CBH} \ee with:
\be
\gamma(\al,\bt)^{\mu}=\al^{\mu}+\bt^{\mu}+c^{\mu}_{\delta
\nu}\al^{\delta}\bt^{\nu}+...
\ee
The relation˜(\ref{CBH}) leads to the associative CBH star
multiplication. The $*$-product defined by this formula is based
on a Weyl map with the exponential of the previous expression:
\be
\Omega_1(f)=\frac{1}{(2\pi)^{n/2}}\int\,d^{n}k\tilde f(k) e^{ik\x}
\ee Define $*_1$ as in~\eqn{stardefweylmap}:
\be
(f*_1 g)(x)=\Omega_1^{-1}\left(\Omega_1(f)\Omega_1(g)\right)
\label{stardef1} \ee Using the CBH formula for the a set of
operators $\hat x_\mu$ satisfying~\eqn{kmincomm}, the infinite
series of nested commutators appearing in the integral can be
integrated explicitly~\cite{KLM2} to give: \bea
e^{ik^\mu\x_\mu}e^{il^\nu\x_\nu}&=&e^{ir(k,l)^\mu\x_\mu}\nn\\
r^0&=&k^0+l^0\\ r^i&=&\frac{\phi(k^0)e^{\lambda
l^0}k^i+\phi(l^0)l^i}{\phi(k^0+l^0)}\\ \phi(a)&=&\frac{1}{a\lambda
}(1-e^{-a\lambda})\label{CBHf} \eea We have that
\be
e^{ik^\mu x_\mu}*_1e^{il^\nu x_\nu}=e^{ir(k,l)^\mu x_\mu}
\label{CBHstar} \ee The product among the generators is obtained
differentiating twice~\eqn{CBHstar} and setting $k=l=0$: \bea
x_0*_1x_0&=&x_0^2\nn\\
x_0*_1x_i&=&x_0x_i+\frac{i\lambda}{2}x_i\nn\\
x_i*_1x_0&=&x_0x_i-\frac{i\lambda}{2}x_i\nn\\ x_i*_1x_i&=&x_i^2
\label{xstar1} \eea of course these reproduce the commutation
rules for $\kappa$-Minkowski space time.

\subsection{The Time Ordered Product \label{sectstar2}}
The time ordered product is a modification of the previous one and
has its roots in the bicrossproduct structure of
$\kappa$-Poincar\'e. It has been first proposed in~\cite{timeord}
and subsequently investigated, for example,
in~\cite{MO,KMLS,KLM2}. One defines the time ordering $:~:$ for
which, in the expansion of a function, all powers of $x_0$ appear
to the right of the $x_i$'s. The relation for a time ordered
exponential is:
\be
e^{i\alpha^\mu\x_\mu}=e^{i\phi(\alpha_0)\alpha^i\x_i}e^{i\alpha^0\x_0}=\,
:\!e^{i(\alpha^0\x_0+\phi(\alpha_0)\alpha^i\x_i)}: \label{timeord}
\ee
where $\phi$ has been defined in˜(\ref{CBHf}). The
relation˜(\ref{timeord}) leads to another associative product via
the map
\be
\Omega_2(f)=\frac{1}{(2\pi)^{n/2}}\int d^{n}k\tilde f(k) :e^{ikx}:
\ee
defined as before by:
\be
(f*_2 g)(x)=\Omega_2^{-1}\left(\Omega_2(f)\Omega_2(g)\right)
\label{stardef2}
\ee
The product of two exponentials is
\be
e^{ik^\mu x_\mu}*_2e^{il^\nu
x_\nu}=e^{i(k^0+l^0)x_0+i\left(k^i+e^{\lambda k^0}l^i\right)x_i}
\label{expstar2} \ee and the product among the generators: \bea
x_0*_2x_0&=&x_0^2\nn\\ x_0*_2x_i&=&x_0x_i\nn\\
x_i*_2x_0&=&x_0x_i-i\lambda x_i\nn\\ x_i*_2x_i&=&x_i^2
\label{xstar2} \eea

\subsection{The Reduced Moyal Product}
This product is a particular case of a general class of
$*$-products for three-dimensional Lie algebras, introduced
in~\cite{selene}, which for the \kmin\ case it can be easily
generalized to an arbitrary number of dimensions. The idea is to
obtain a product in an $n$ dimensional space by considering it as
a subspace of an higher dimensional symplectic phase space,
equipped with the the usual Poisson bracket, and a Moyal $\star$
product (with deformation parameter $\lambda$). The product is
then defined by first lifting the functions from the smaller space
to the phase space, multiplying them in the higher space, and then
projecting back to the smaller space. The form of the lift (a
generalized Jordan-Schwinger map) and the canonical structure of
the Moyal product ensure that this procedure defines a good $*$
product in the smaller dimensions.

We indicate the coordinate on the phase space $\real^6$ with the
notation: $\mathbb{R}^6 \ni u=(q_1,q_2,q_3;p_1,p_2,p_3)$. We need
to define a map $\pi$:
\be
\pi: \mathbb{R}^6\rightarrow \mathbb{R}^4
\ee
or equivalently the map $\pi^*$ which pulls smooth functions on
$\real^4$ to smooth functions on $\real^6$. The map $\pi$ is
explicitly realized with four functions of the $p$'s and $q$'s,
which we call $x_\mu$. The requirement is that the six dimensional
Poisson bracket of $x$'s reproduce the ``classical'' \kmin\
algebra\footnote{This algebra is an extension of the three
dimensional Lie-algebra $sb(2,\complex)$ of $2\times 2$ triangular
complex matrices with zero trace treated in~\cite{selene}.}:
\be
\{x_0,x_i\}=  x_i,\;\;\;\{x_i,x_j\}=0\;\;\; i,j=1,2,3
\ee
The $*_3$ product is then defined by
\be
\pi^*(f*_3g)=\pi^*f\star\pi^*g
\ee
The fact that, after performing the (nonlocal) product in six
dimensions, we are left with a function still defined using only
the four dimensional coordinates is ensured by the existence of
two local functions, $H_1$ and $H_2$, with the property that
\be
L_{H_i}\left(\pi^*f\right)=0 \ee and
\be
L_{H_i}\pi^*\left(f(x)\star g(x)\right)=0
\ee
In other words, from the six dimensional point of view, the $x$'s
commute with the $H$'s, and this commutation is stable under the
$\star$-product, thus ensuring that if we multiply a function only
of the $x$'s, the product does not depend on the extra
coordinates. Upon the identification of the parameter $\theta$
with $\lambda$ we obtain a \kmin\ product on $\real^4$.

One of the possible realization of $\pi$ is:
\bea
x_0&=&- \sum_iq_ip_i\nn\\
x_i&=&q_i\label{jsmap}
\eea
Two independent commuting functions are:
\bea
H_1&=&\arctan\frac{q_2}{q_3}\nn\\&~&\nn\\
H_2&=&\arctan\frac{q_3}{q_1}
\eea
Note however that the representation ~\eqn{jsmap} is not unique.
For example the following choice works as well:
\bea
x'_0&=&- \sum_ip_i\nn\\
x'_i&=&e^{q_i} \label{seleneprod'}
\eea
with the commuting functions
\bea
H_1'&=&e^{q_2-q_3}\nn\\
H_2'&=&e^{q_3-q_1}
\eea
connected by a singular canonical transformation to the previous
one.

We will denote the products with the choices~\eqn{jsmap}
and~\eqn{seleneprod'} as~$*_3$ and $*_4$ respectively. The
explicit products between the generators are in~\cite{selene}:
\bea
x_0*_3x_0&=&x_0^2+\frac{3}{4}\lambda^2\nn\\
x_0*_3x_i&=&x_0x_i+i\frac\lambda2x_i\nn\\
x_i*_3x_0&=&x_0x_i-i\frac\lambda2x_i\nn\\ x_i*_3x_i&=&x_i^2
\eea
and
\bea
x_0*_4x_0&=&x_0^2\nn\\
x_0*_4x_i&=&x_0x_i+i\frac\lambda2x_i\nn\\
x_i*_4x_0&=&x_0x_i-i\frac\lambda2x_i\nn\\
x_i*_4x_i&=&x_i^2
\eea
It may be noticed that these relations are the same as the one for $*_1$
in~\eqn{xstar1}. The two products however, although similar, do not
coincide for generic functions.

\section{Weyl Systems for \kmin}
In this section we will show how the various products presented
earlier are particular instances of generalized Weyl systems. The
starting point for the construction of the product is the
identification of the $W(k)$'s, which enables the calculation of
the particular relation~\eqn{genweylrel} for the various cases. In
other words we will give an explicit realization of the group
composition law $\oplus$. All Weyl systems presented here have
$\omega=0$, and in all cases a straightforward calculation
verifies relation~(\ref{3}).
\subsection{Weyl System for the CBH product}
In this case the relation is given by the CBH product, which for
\kmin\ has been given in eqs.~(\ref{CBH}-\ref{CBHf}). The
composition $\oplus_1$ can be calculated to give:
\bea
(k\oplus_1 k')^0&=&k^0+{k'}^0\nn\\ (k\oplus_1
k')^i&=&\frac{\phi(k^0)k^i+e^{\lambda
k^0}\phi({k'}^0){k'}^i}{\phi(k^0+{k'}^0)}
\eea
where $\phi$ has been defined in eq.~\eqn{CBHf}. We have that
\be
e^{ik\hat{x}}=\Omega\left(e^{ikx}\right) \ee There is one check
which has to be performed to ensure that $\oplus_1$ defines a
group. Relation~\eqn{ominus} is verified with the definition
\be
\bar{k}=(-k^0,-k^i)
\ee
in fact
\be
\overline{(k\oplus_1 k')}=\bar{k'}\oplus\bar{k}
\ee

\subsection{Weyl System for the Time Ordered Product}
This case is similar to the previous one, and a direct calculation
using the CBH relations for the time ordered exponentials give:
\bea
(k\oplus_2k')^0&=&k^0+{k'}^0\nn\\
(k\oplus_2k')^i&=&k^i+e^{\lambda k^0}{k'}^i
\eea
with
\be
\bar{k}=(-k^0,-e^{-\lambda k^0}k^i)
\ee
This momenta composition reflects the coproduct in the
bicrossproduct basis, thus rendering the time ordered exponential
a natural basis for the space of functions in this case. Note that
\be
:\!e^{ik\hat{x}}:=\Omega_{1}\left(e^{ik^ix_i}*_1e^{ik^0x_0}\right)\label{newcar}
\ee So the product $*_2$ can be expressed by the choice of a
different $W$ \emph{of the $*_1$ product}.

\subsection{Weyl Systems for the Reduced Moyal products}
The path to the definition of the reduced products is
intrinsically different from the first two. These are defined
using straightforwardly the CBH formula with a specific ordering.
The reduced product instead comes from a four dimensional
reduction of a six dimensional product. There is therefore no
warranty that it is possible to obtain them form a \emph{four}
dimensional Weyl system. This is nevertheless possible.

Notice that if we were to define $W$ as the $*_3$ or $*_4$
exponential of $ik^\mu x_\mu$ we would find the CBH product. This
is because the $*$ commutator of the $x$'s are all the same. We
must therefore use another quantity, and we could use the ordinary
exponential. Care must be taken however because this is not an
unitary operator (with the hermitean conjugation defined in
eq.~\eqn{conjuga}). So that it is necessary to normalize it. The
calculations of the product of two exponentials are in Appendices
A and B. We define $W_3$ as
\be
W_3\left(k\right)=|a(k^0,-k^0)|^{3/2}\Omega_{3}\left(e^{ikx}\right)
\label{W3def} \ee with \bea
a(k^0,l^0)&=&1+\frac{\lambda^2}{4}k^0l^0\nn\\
b(k^0)&=&1+\frac{\lambda}{2}k^0 \label{defab} \eea {}From
equation~\eqn{new} we can read \be (k\oplus_3
k')^\mu=k^\mu+{k^\mu}' -\frac{\lambda}{2a}( {k'}^0 b(k^0)
k^\mu-{k'}^0 b(-{k'}^0) {k'}^\mu) \ee with $\bar k=-k$. Unlike all
other composition laws, $\oplus_3$ is the only one in which the
``time-like'' coordinate has a nonabelian structure:\footnote{Note
also that $\oplus_{3}$ is not well defined for all
$\left(k,k^{\prime}\right)$, so that we do not have a group. The
integral (\ref{deformedproduct}) is anyway well
defined.}$(k\oplus_3 k')^0\neq(k'\oplus_3 k)^0$.

For the $*_4$ case, the exponential is unitary, and it is possible
to define $W_4(k)=e^{ikx}$. In Appendix B we calculate the
composition rule $\oplus_4$ which results:
\be
k\oplus_4 k'=(k^0+{k'}^0,e^{-\frac{\lambda}{2}
{k'}^0}\vk+e^{\frac{\lambda}{2} k^0}\vk')\label{oplus4}
\ee
with $\bar k=-k$.

Notice that, while the composition law $\oplus_2$ is connected to
the coproduct~\eqn{coproducts} in the bicrossproduct basis, the
$\oplus_4$ is related to the coproduct in the standard
basis~\cite{LNRT}:
\bea
\Delta P_0&=& P_0\otimes 1 + 1 \otimes P_0\nn\\
 \Delta M_i&=& M_i\otimes 1 +1 \otimes M_i\nn\\
 \Delta P_i &=& P_i\otimes e^{-\frac{\lambda P_0}{2}} +
 P_i\otimes e^{\frac{\lambda P_0}{2}}\nn\\
  \Delta N_i
&=&  N_i\otimes e^{\frac{\lambda P_0}{2}}+e^{-\frac{\lambda
P_0}{2}}\otimes
N_i+\frac{\lambda}{2}\varepsilon_{ijk}\left(P_j\otimes
M_ke^{\frac{\lambda P_0}{2}}+e^{-\frac{\lambda P_0}{2}}M_j\otimes
P_k\right) \label{coproductssta}
\eea
and in particular to the symmetric coproduct for $P_i$. The
similarity goes beyond it, as the $*_4$ product can be written as
a ''symmetrically ordered'' product with respect to the $x_0$
coordinate defining:
\be
\ddag e^{ik\hat{x}}\ddag = e^{\frac{ik^0}{2}\hat{x}_{0}}e^{-i
k^{i}\hat{x}_{i}}e^{\frac{ik^0}{2}\hat{x}_{0}} \ee so that
\be
\ddag e^{ik\hat{x}}\ddag \,\ddag e^{il\hat{x}}\ddag = \ddag
e^{i(k\oplus_4 l)\hat{x}}\ddag \ee In analogy with (\ref{newcar})
we have \be W_{4}\left(k\right)=\Omega_{1}\left(
e^{\frac{ik^0}{2}x_{0}}*_{1}e^{-i
k^{i}x_{i}}*_{1}e^{\frac{ik^0}{2}x_{0}}\right)\ee

\section{Conclusions}
A noncommutative geometry is defined by a $*$-algebra, and
deformations of the algebras of functions on a manifold are
particularly important examples. In this paper we have shown that
it is in principle possible to obtain noncommutative products
between functions which generalize the commutation relations of
Lie algebras, possibly with central extensions. This has been done
generalizing Weyl systems by the substitution of the usual abelian
sum with a nonabelian composition law.

We have seen that four different $*$-products which generalize the
commutation relations of \kmin\ fit nicely in this framework. This
enabled us to perform a comparative study of these products, and
to discover several relations among them, and with the coproducts
and other structures which characterize the quantum algebra
$\kappa$-Poincar\'e.  These consideration will be helpful for the
ultimate goal of the construction of field theories on these
noncommutative spaces.

\subsection*{Acknowledgments}
We thank G.~Amelino-Camelia, J.M.~Gracia-Bond\'{\i}a, G.~Marmo and J. C.
V\'{a}rilly for helpful discussions and correspondence. The work of F.L.\
is supported in part by the {\sl Progetto di Ricerca di Interesse Nazionale
{\em SInteSi 2000}}.

\setcounter{section}{0}
\appendix{Calculation of the $*_3$ product of two exponentials}
In this appendix we give some details of the calculations of the
product of two (ordinary) exponential functions. We will use the
following (equivalent) form of the product~\eqn{eq:Moyal} for the
product of two functions $f,g$ on $\real^{6}$:
\be
f(u)\star g(u)=(2\pi)^{-6}\int d^6sd^6t
f(u+\frac{\lambda}{2}J_{6}s)g(u+t)e^{-ist} \label{moyal}
\ee
where $u=(q_1,...p_{3})$ and $J$ denotes the antisymmetric matrix:
$$
{J_{6}}=\left(
\begin{array}{ccc}
0& &I_3\\
-I_3& &0
\end{array}
\right)
$$
Let us express $u=(\vec{q},\vec{p})$ and introduce the following
notation:
\bea
u&=&(\vq,\vp)\nn\\
s&=&(\vs_1,\vs_2)\nn\\
t&=&(\vt_1,\vt_2)
\eea
in which all vectors belong to $\mathbb{R}_3$. Using the integral
form for a Moyal product~(\ref{moyal}), the deformed (six
dimensional) product of two exponential is:
\be
e^{ik^{\mu}x_{\mu}}\star e^{il^{\mu}x_{\mu}}=(2\pi)^{-6} \int
d^6sd^6t e^{ik^{\mu}x_{\mu}(u+\frac{\lambda}{2}
J_6s)}e^{il^{\mu}x_{\mu}(u+t)} e^{-ist}
\ee
with (using~\eqn{jsmap} for the last step):
\be
ik^{\mu}x_{\mu}(u)\equiv ikx=ik^0x_0-i\vk\cdot\vx=-i(k^0\vq
\cdot\vp+\vk \cdot\vq)
\ee
the arguments of the $x$'s become
\be
u+\frac{\lambda}{2}J_6s=(\vq+\frac{\lambda}{2}\vs_2,\vp-\frac{\lambda}{2}\vs_1)
\ee
\bea
ik^{\mu}x_{\mu}(u+\frac{\lambda}{2}J_6s)
&=&-i[k^0(\vec{q}+\frac{\lambda}{2}\vec{s_2})
(\vec{p}-\frac{\lambda}{2}\vec{s_1})+\vec{k}\cdot(\vec{q}+\frac{\lambda}{2}\vec{s_2})]\nonumber \\
&=&-i(k^0\vq\cdot\vp+\vk\cdot\vq)-i\frac{\lambda}{2}(k^0(\vs_2\cdot\vp-\vq\cdot\vs_1)+\vk\cdot\vs_2)+
i\frac{\lambda^2}{4}k^0\vs_1\cdot\vs_2 \nonumber \\
&=&ikx-i\frac{\lambda}{2}(k^0(\vs_2\cdot\vp-\vq\cdot\vs_1)+\vk\cdot\vs_2)+i\frac{\lambda^2}{4}k^0\vs_1\cdot\vs_2
\eea
\bea
il^{\mu}x^{\mu}(u+t)&=&-i[l^0(\vec{q}+\vt_1)(\vp+\vec{t_2})+\vec{l}\cdot(\vec{q}+\vec{t_1})]\nn\\
&=&ilx-i(l^0(\vq\cdot\vt_2+\vp\cdot\vt_1+\vt_1\cdot\vt_2)+\vl\cdot\vt_1)
\eea
and
\bea
e^{ik^{\mu}x_{\mu}}&\star
&e^{il^{\mu}x_{\mu}}=(2\pi)^{-6}e^{i(k+l)x} \int
d\vec{s}_1d\vec{s}_2d\vec{t}_1d\vec{t}_2
e^{-i\frac{\lambda}{2}(k^0(\vs_2\cdot\vp-\vq\cdot\vs_1)+\vk\cdot\vs_2)+i\frac{\lambda^2}{4}k^0\vs_1\cdot\vs_2
}\nn\\
&&e^{-i(l^0(\vq\cdot\vt_2+\vp\cdot\vt_1+\vt_1\cdot\vt_2)+\vl\cdot\vt_1)}
e^{-i(\vec{s}_1\cdot\vec{t}_1+\vec{s}_2\cdot\vec{t}_2)}\eea
Reordering the exponentials:
\bea
e^{ik^{\mu}x_{\mu}}\star
e^{il^{\mu}x_{\mu}}&=&(2\pi)^{-6}e^{i(k+l)x}
\int d\vec{s}_1d\vec{s}_2d\vec{t}_1d\vec{t}_2e^{i\vs_1(\frac{\lambda}{2}k^0\vq+\frac{\lambda^2}{4}k^0\vs_2-\vt_1)}\nn\\
&&e^{-i\vt_2(l^0\vt_1+l^0\vq+\vs_2)}
e^{-i\frac{\lambda}{2}\vs_2(k^0\vp+\vk)-i\vt_1\cdot(\vl+l^0\vp)}
\eea
at this point we can make the integration in the $\vs_1$ e $\vt_2$
variables:
\bea
\frac{1}{(2\pi)^3}\int d\vs_1
e^{i\vs_1(\frac{\lambda}{2}k^0\vq+\frac{\lambda^2}{4}k^0\vs_2-\vt_1)}&=&
\delta^{(3)}(\frac{\lambda}{2}k^0\vq+\frac{\lambda^2}{4}k^0\vs_2-\vt_1)\nn\\
\frac{1}{(2\pi)^3}\int d\vt_2
e^{-i\vt_2(l^0\vq+l^0\vt_1+\vs_2)}&=&\delta^{(3)}(l^0\vq+l^0\vt_1+\vs_2)
\eea
to obtain:
\bea
e^{ik^{\mu}x_{\mu}}\star e^{il^{\mu}x_{\mu}} &=&e^{i(k+l)x}\int
d\vs_2d\vt_1\delta^{(3)}
(\frac{\lambda}{2}k^0\vq+\frac{\lambda^2}{4}k^0\vs_2-\vt_1)\delta^{(3)}(l^0\vq+l^0\vt_1+\vs_2)\nonumber \\
&&e^{-i\frac{\lambda}{2}\vs_2(k^0\vp+\vk)-i\vt_1\cdot(\vl+l^0\vp)}
\eea
and making integral in  $d\vt_1$ we have:
\bea
e^{ik^{\mu}x_{\mu}}\star e^{il^{\mu}x_{\mu}} &=&e^{i(k+l)x}\int
d\vs_2
e^{-i\frac{\lambda}{2}\vs_2\cdot(k^0\vp+\vk)-i\frac{\lambda^2}{4}k^0\vs_2(\vl+l^0\vp)
-i\frac{\lambda}{2}k^0\vq(\vl+l^0\vp)}\nn\\
&&\delta^{(3)}(\vs_2(1+\frac{\lambda^2}{4}k^0l^0)+l^0(1+\frac{\lambda}{2}k^0)\vq)
\eea
Using $a$ and $b$ defined in~\eqn{defab}, we make the last
integration to obtain:
\be
\delta(\vs_2a(k^0,l^0)+l^0b(k^0)\vq)=
\frac{1}{|a(k^0,l^0)|^3}\delta(\vs_2+l^0\frac{b(k^0)}{a(k^0,l^0)}\vq)
\ee
which fixes $\vs_2=-l^0\frac{b}{a}\vq$, so that the integral becomes:
\bea
&&\frac{1}{|a(k^0,l^0)|^3}e^{i\frac{\lambda}{2}l^0\frac{b}{a}
\vq\cdot(k^0\vp+\vk)+i\frac{\lambda^2}{4}k^0l^0\frac{b}{a}\vq(\vl+l^0\vp)-
i\frac{\lambda}{2}k^0\vq(\vl+l^0\vp)}\nn\\
&=&\frac{1}{|a(k^0,l^0)|^3}e^{i\frac{\lambda}{2}l^0\frac{b}{a}
(-k^0x_0+\vk\cdot\vx)+i\frac{\lambda^2}{4}k^0l^0\frac{b}{a}(\vl\cdot\vx-l^0x_0)-
i\frac{\lambda}{2}k^0(\vl\cdot\vx-l^0x_0)}\nn\\
&=&\frac{1}{|a(k^0,l^0)|^3}e^{-i\frac{\lambda}{2}l^0\frac{b}{a}
kx-i\frac{\lambda^2}{4}k^0l^0\frac{b}{a}lx+i\frac{\lambda}{2}k^0lx}\nn\\
&=&\frac{1}{|a(k^0,l^0)|^3}e^{-i\frac{\lambda}{2a}
(l^0bk+\frac{\lambda}{2}k^0l^0bl-\frac{\lambda}{2}k^0la)x}\nn\\
&=&\frac{1}{|a(k^0,l^0)|^3}e^{-i\frac{\lambda}{2a}(l^0b(k^0)k-k^0b(-l^0)l)x}
\eea
The final result is:
\be
e^{ikx}\star
e^{ilx}=\frac{1}{|a(k^0,l^0)|^3}e^{i(k+l)x}e^{-i\frac{\lambda}{2a(k^0,l^0)}
(l^0b(k^0)k-k^0b(-l^0)l)x}
=\frac{1}{|a(k^0,l^0)|^3}e^{i(k\oplus_3 l)x}\label{new}
\ee
It results also $\bar k=-k$. From relation~\eqn{new} can be seen
that the function $e^{ikx}$ is not unitary for the product~$*_3$,
and to make it unitary one should renormalize it dividing by
$|a(k^0,k^0)|^{3/2}$, thus finding~\eqn{W3def}.

\appendix{Calculation of the $*_4$ product of two exponentials \label{appstar4}}
In this second appendix we consider the map given by
relation~\eqn{seleneprod'}, which gives rise to the product~$*_4$.
Again with the use of~\eqn{moyal} we calculate the product among
exponential functions:
\be
f(u)=e^{ikx},\ \ \ g(u)=e^{ilx}
\ee
where $x=x(u)$ and $u,s,t$ are defined as in the previous
appendix. We have then:
\bea
ikx(u+\frac{1}{2}\lambda
Js)&=&-i\left(k^0\sum_{i=1}^3(p_i-\frac{\lambda}{2} s_{1i})+
\sum_{i=1}^3k_ie^{q_i+\frac{\lambda}{2} s_{2_i}}\right)\nn\\
&=&k^0x_0+i\sum_{i=1}^3\left(\frac{\lambda}{2}k^0s_{1i}-k_ix_ie^{\frac{\lambda}{2}s_ {2i}}\right)\\
ilx(u+t)&=&-i\sum_{i=1}^3\left(l^0(p_i+t_{2i})+l_ie^{q_i+t_{1i}}\right)\nn\\
&=&l^0x_0-i\sum_{i=1}^3[l^0t_{2i}+l_ix_ie^{t_{1i}}]
\eea
Performing the integral:
\bea
e^{ikx}*_4e^{ilx}=&e^{i(k^0+l^0)x_0}&\int ds dt
e^{i\sum_i\left(\frac{\lambda}{2}k^0s_{1i}-k_ix_i
e^{\frac{\lambda}{2}s_{2i}}\right)}
e^{i\sum_i^3\left(-l^0t_{2i}-l_ix_ie^{t_{1i}}\right)}e^{-i\sum_i^3(s_{1i}t_{1i}+s_{2i}t_{2i})}\nonumber\\
=&e^{i(k^0+l^0)x_0}&\prod_i^3\int ds_{2_i}dt_{1_i}\delta(\frac{\lambda}{2}k^0-t_{1i})
\delta(l^0+s_{2i})e^{-ik_ix_ie^{\frac{\lambda}{2}s_{2i}}}e^{-il_ix_ie^{t_{1i}}}\nn\\
=&e^{i(k^0+l^0)x_0}&\prod_i^3[e^{-ik_ix_ie^{-\frac{\lambda}{2}l^0}}e^{-il_ix_ie^{\frac{\lambda}{2}k^0}}]\nn\\
=&e^{i(k^0+l^0)x_0}&[e^{-i\vk\vx e^{-\frac{\lambda}{2}l^0}}e^{-i\vl\vx e^{\frac{\lambda}{2}k^0}}]\nn\\
=&e^{i(k\oplus_4 l)x}&
\eea

\end{document}